\documentclass[traditabstract]{aa} 
\usepackage{txfonts,psfig}

\begin{document}

\title{The important role of evolution in the 
Planck $Y_{SZ}$--mass calibration}
\titlerunning{Evolution of $Y_{SZ}$-mass scaling} 
\author{S. Andreon}
\authorrunning{S. Andreon}
\institute{
INAF--Osservatorio Astronomico di Brera, via Brera 28, 20121, Milano, Italy,
\email{stefano.andreon@brera.inaf.it} \\
}
\date{Accepted ... Received ...}
\abstract{In light of the tension between cosmological parameters
from Planck cosmic
microwave background and galaxy clusters, we revised the Planck
analysis of the $Y_{SZ}$-mass calibration to allow evolution
to be determined by the data instead of being
imposed as an external constraint.
Our analysis uses the very same data and Malmquist bias
corrections as used by the Planck team 
in order to emphasize that differences in
the results come from differences in the assumptions. 
The evolution derived from 71 calibrating clusters, 
with $0.05<z<0.45$, is proportional
to $E^{2.5\pm0.4}(z)$, so inconsistent with the self--similar evolution
($E^{2/3}$) assumed by previous analyses. When
allowing for evolution, the slope of $Y_{SZ}$--mass
relation turns out to be $1.51\pm0.07$, which is shallower by $4.8\sigma$
than the value derived when assuming self-similar evolution, introducing
a mass--dependent bias. The non--self--similar
evolution of $Y_{SZ}$ has to be accounted for in analyses
aimed to establish the biases of Planck masses. 
}
\keywords{  
Galaxies: clusters: general  --- Methods: statistical --- 
Cosmology: observations ---  Galaxies: clusters: intracluster medium
}

\maketitle

\section{Introduction}

The Sunyaev-Zel'dovich (SZ) effect, which is the distortion caused by
the high energy electrons of the intracluster medium on the cosmic
microwave background (CMB) photons, has been used by
Ade et al. (2014a) to put constraints on
cosmological parameters using Planck data.
The cosmological constraint comes from matching the galaxy 
cluster abundance (per unit observable) to
the cluster mass function (per unit mass), via a 
mass--observable relation. At present, the Planck mass--observable 
calibration is a two--step process (Ade et al. 2014a,b):
first, the $Y_X$ mass proxy ($Y_X = T_X M_{gas}$,
Kravtsov et al. 2006) is
calibrated against hydrostatic masses (Arnaud et al. 2010), and,
second, this calibration is transferred to the $Y_{SZ}$ proxy using the
measured $Y_{SZ}$ values for 71 clusters (with $Y_X$ values) 
observed by Planck (Ade et al. 2014a,b). This calibration 
uses clusters spread over a sizeable redshift range ($0<z<0.5$, see
Fig.~1) 
and assumes
a given (self-similar) evolution for the mass proxies. The assumption
of a self-similar evolution relies on the
the self-similar model (Kaiser 1991) and numerical
simulations (Kravtsov et al. 2006) and lacks an observational
determination, i.e. a calibration done using
directly measured masses (i.e. weak lensing) at various redshifts. 

The cosmological constraints derived by Planck using galaxy
clusters differ from those derived (mostly) from the CMB (Ade et al. 2014a). 
In particular, the best fit CMB cosmology predicts many more
clusters than observed, suggesting that
the current Planck-estimated masses may be underestimated.

Rozo et al. (2012) use
Chandra data from Vikhlinin et al. (2009) and a subsample of
an early Planck sample
(Ade et al. 2011) to argue
that the calibration may have an intercept and, possibly, 
a slope, lower than derived in Ade et al. (2011), which uses the same
$Y_{SZ}$ values but $Y_X$ values derived from XMM data. 
The lower intercept
agrees with X--ray expectations (Rozo et al. 2012).
Rozo et al. (2012) also
find no evidence of any evolution in
the intercept of the $Y_{SZ}-Y_{X}$ relation,
giving indirect support
to the self--similar evolution assumed by the Planck team but
under a number of restrictive hypothesis. For example, they
hold fixed the slope of the relation and they suppose to perfectly
know the evolution of $Y_{X}$.

Recently, von der Linden et al. (2014) have compared Planck SZ mass estimates
to weak--lensing masses and found that Planck
masses need to be underestimated at a $\sim 1.6\sigma$ level and, 
with modest evidence, need it even more so 
with increasing masses. Similar 
clues
have also been found by 
Sereno et al. (2014) for 
the von der Linden et al. (2014) sample 
and for other samples as well.
Analyses of these two works assume a self--similar evolution
and therefore attribute the tilted scaling to a mass bias. 
However, Andreon \& Congdon (2014) and Gruen et al. (2014) 
note that the mass slope and the evolution of the scaling may be
collinear (degenerate)\footnote{ Sereno et al. (2014) also mention this
possibility, but their analysis does not account for it.}. Therefore,  
the claimed mass bias can be
a manifestation of an evolution in the $Y_{SZ}$-mass 
scaling.

Because the evolution of $Y_{SZ}$ is unlikely to be perfectly known 
and because there are hints in the literature of possible mass--dependent
biases that could instead be manifestations of redshift effects,
in this paper we 
independent assess the Planck mass calibration. Our approach 
is free of
evolutionary assumptions and breaks the mass--redshift degeneracy.
More specifically, we fit the $Y_{SZ}$--mass 
allowing (solving for) evolutionary effects
that have been neglected in the previous mass calibration. If ignored,
these effects lead
to a mass--dependent bias. In particular, we 
use the very same input data as used by the Planck team, 
in order to emphasize that differences in
the results come from differences in the assumptions. 

\begin{figure}
\centerline{
\psfig{figure=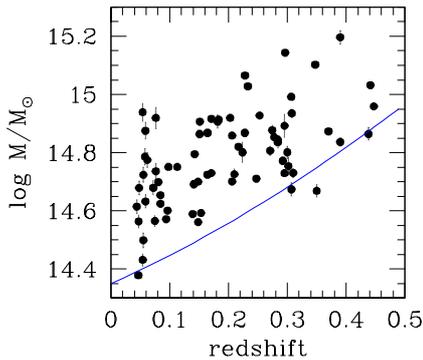,width=6truecm,clip=}
}
\caption[h]{Mass vs redshift plot of the Planck 
calibrating sample. Points are $M^{YX}_{500}$ masses.
The solid line marks the adopted 
threshold mass $M_{thr}$ of our secondary analysis (sec.~2.1).
}
\end{figure}

We assume $\Omega_M=0.3$, 
$\Omega_\Lambda=0.7$, and $H_0=70$ km s$^{-1}$ Mpc$^{-1}$. 
Results of stochastic computations are given
in the form $x\pm y$ where $x$ and $y$ are 
the posterior mean and standard deviation. The latter also
corresponds to 68 \% intervals, because we only summarized
posteriors close to a Gaussian in that way. All $\log$ are in base
10.

\section{Revisiting the $Y_{SZ}$ calibration}

Our starting point is the catalogue of the $Y_X$-derived masses,
$M^{YX}_{500}$, and Malmquist--bias--corrected $Y_{SZ}$ values 
delivered by the Planck team for the 71 calibrating
clusters\footnote{The table
is available at http://szcluster-db.ias.u-psud.fr.}. 
Their calibration sample has an increasing limiting
mass with redshift (Fig.~1). 

\begin{figure*}
\centerline{
\psfig{figure=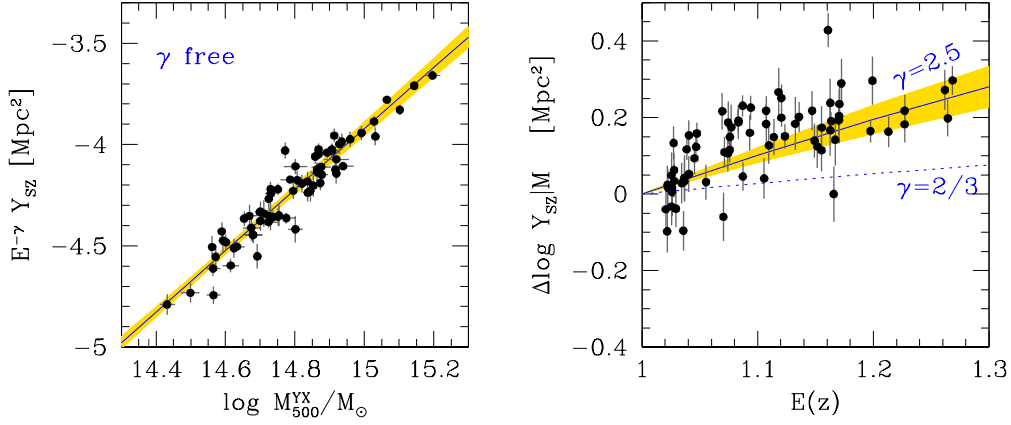,height=6truecm,clip=}
}
\caption[h]{Mass-$Y_{SZ}$ scaling (left--hand panel) and residuals (observed minus
expected) as a function of redshift (right--hand panel).
The solid line marks the mean fitted regression line. 
The shaded  region marks the 68\% uncertainty
(highest posterior density interval) for the regression. In the left--hand panel,
measurements are corrected for evolution. In the right--hand panel 
the dotted line shows
the $E^{2/3}$ dependency assumed by the Planck team.
}
\end{figure*}

Following Ade et al. (2014a,b),  
we fit the data with the function
\begin{equation}
\log Y_{SZ,z}=\log Y_{SZ,z=0} + \gamma \log E(z) + s (\log (M^{YX}_{500}/M_{ref}))
\end{equation}
allowing a log--normal
scatter in $Y_{SZ}|M^{YX}_{500}$ around the mean relation, 
where $M_{ref}=6\ 10^{14} M_{\odot}$. 
In contrast to Ade et al. (2014a,b), we leave the $\gamma$ parameter free, 
rather than freezing it to the self--similar value ($2/3$). This is an important
difference, because we no longer assume the self--similar evolution,
and we are trying to solve for the $Y_{SZ}$-mass scaling {\it and}
its evolution. The fit is performed using 
the standard errors--in--variables
Bayesian regression (Dellaportas \& Stephens, 1995; 
see Andreon \& Hurn 2013 for an introduction)\footnote{The fitting code
is available at \hfill \break
http://www.brera.mi.astro.it/$\sim$andreon/fitSZplanck.bug.}.

The mass-$Y_{SZ}$-redshift fit results
are shown in Fig.~2. We found 
that $\log Y_{SZ}$ scales with $\log M^{YX}_{500}$  
with slope $s=1.51\pm0.07$, with a negligible intrinsic scatter 
($\log Y_{SZ}|M^{YX}_{500} =0.06\pm0.01$ dex),
\begin{eqnarray}
\log Y_{SZ,z}  &=  (1.51\pm0.07) (\log (M^{YX}_{500}/M_{ref})) -4.26\pm0.02 + \nonumber \\
\quad &\quad +(2.5\pm0.4) \log E(z) \hfill
\end{eqnarray}
with some covariance between the mass slope $s$ and the
evolutionary term $\gamma$, as shown in Fig.~3.
The evolution of the $Y_{SZ}$-mass scaling is well constrained
by the data, $\gamma=2.5\pm0.4$,
mainly because a sizeable range of masses are available
at a given redshift (because of the vertical spread at a fixed
$z$ in Fig.~1). Indeed, if the relation in Fig.~1 were 
scatterless, then a complete collinearity (degeneracy)
between the $Y_{SZ}$-mass slope and the redshift evolution would be
present, meaning
the slope evolution could not be measured without assuming a slope
for the relation. The
scatter breaks the collinearity. This is not
the case for the Gruen et al. (2014) sample, formed by just seven
clusters, as also noted by the authors.

The evolutionary term $\gamma$ derived from the data, $\gamma=2.5\pm0.4$, is 
much larger than the self--similar value assumed
in the Ade et al. (2014a,b) fitting, $\gamma=2/3$. 
Indeed, $\gamma=2/3$ is rejected by the data
because the posterior probability of $\gamma$ = $2/3$ is extremely
low (i.e. more than $>500$ times smaller than the modal value). Furthermore, 
$\gamma=2/3$ is just outside the boundary of
the 99.9 \% interval of $\gamma$ (Fig.~3), and only 0.01\% of the Monte Carlo samples
have lower $\gamma$. Expressed in another way, 
the $\gamma$ value derived from the data differs from the one
assumed in the Ade et al. (2014a,b) fit in a way that makes 
the new and old mass slopes different by
$4.8$ times the slope error quoted in Ade et al. (2014a,b).
The disagreement between the evolution pointed
out by the data and the one
assumed in previous works (e.g. Rozo et al. 2012, Sereno et al. 2014,
von der Linden et al. 2014, Ade et al. 2014a,b)
is also illustrated in the right--hand
panel of Fig.~2: $Y_{SZ}$ residuals (i.e. observed minus fitted) have a slope
of $\sim 2.5$ (solid line), not $2/3$ (dotted line).

Equation 3 has an immediate consequence: 
the mass slope of quantities like $E^{-2/3} Y_{SZ}$ or any quantity
built from it like Planck masses, 
depends on how the studied clusters are distributed 
in the mass--redshift plane. If clusters
are all at the same redshift, then the correct slope is recovered (because
in such a case $E(z)$ is a constant), but
if the mass distribution changes with redshift (e.g. because of selection
effects), a biased slope is found because of the mismatch between
the true data dependency, $E^{2.5}$, and the assumed evolution, $E^{2/3}$.
The sample used in the $Y_{SZ}$--mass analysis of Rozo et al.
(2012) has a very narrow redshift distribution:
70\% of their sample is at $0.04<z<0.10$. Their slope is, therefore,
unaffected by holding the evolutionary term fixed to $E^{2/3}$. Instead,
the samples used in the $Y_{SZ}$--mass analyses in 
von der Linden et al. (2014) and Sereno et al. (2014)
are distributed over a sizeable redshift range. For
these studies it still needs to be evaluated whether the effect interpreted
as a mass bias is instead a consequence of the assumption of
a self--similar evolution. 

In Figure 4, we show the ratio between the recalibrated and original
masses for the whole Planck cosmological sample and for two redshift
subsamples. At a fixed redshift, points are aligned on a line of
slope $0.28$ (the difference in slope between the two calibrations), 
introducing a mass--dependent bias. The
intercept decreases with increasing redshift or, equivalently,
the abscissa (mass) at which the two calibrations equal each other increases
with redshift. At a
fixed redshift, the most massive clusters tend to have recalibrated masses
five to ten percent higher
than originally quoted (Figure 4), in the same direction
as suggested by
von der Linden et al. (2014) and Sereno et al. (2014), by 
analyses that assume, however, a self--similar evolution. Instead,
the originally derived Planck masses
are biased high at the lowest masses entering in the cosmological sample
(Figure 4). 
Therefore, at face value, our new calibration implies more
clusters than the original calibration, which already produces
more clusters than observed for the Planck CMB cosmology. In other terms, 
the recalibration does not seem to solve the tension between 
the CMB-- and cluster--based cosmologies. 
Nevertheless, we want to emphasize that
$Y_{SZ}$ proxy values of the cosmological sample are derived
by the Planck team when assuming their $Y_{SZ}$-mass scaling, not
ours. 
Therefore our
conclusion about the role of evolution in solving the tension 
between 
the CMB-- and cluster--based cosmologies should be re--evaluated
after consistently deriving $Y_{SZ}$ proxy values 
with the updated calibration (i.e. with eq.~2).
A re--extraction of the $Y_{SZ}$ values
of the Planck cluster catalogue is beyond the scope of this work.
 
Finally, we would like to emphasize that 
the current Planck calibration is indirect because it is based on the
mass surrogate $Y_X$, assumed to be self--similar evolving. Therefore,
the found non--self--similar evolution of the $Y_{SZ}$--mass calibration
can be genuine, i.e. related to
a non--self--similar evolution of the SZ mass proxy itself, or
induced by a possible non--self--similar evolution of the mass surrogate $Y_X$. 
In fact, the evolution
of the $Y_X$--mass scaling lacks an observational determination,
since our current knowledge is currently limited to a consistency 
check (Israel et al.  2014), at the differences of the richness, that has
a directly measured (non--) evolution (Andreon \& Congdon, 2014).

\subsection{Alternative treatment of the mass and selection function}

We now show that the precise schema used
to correct for the sample selection function and for
the Malmquist bias has little practical
consequences. To correct for the Malmquist bias, we now take the
non--Malmquist bias--corrected data and  compute the mass
function and its evolution from the Multidark simulation 
(Prada et al. 2012; the data are available in CosmoSim, Riebe et al. 2013).
We assume that the the selection function is a 
step function with 
a threshold mass, $M_{thr}$, linearly increasing with $E(z)$:
\begin{equation}
\log M_{thr}=2 \ (E(z)-1.1)+14.55 \, 
\end{equation}
as shown in Fig.~1.
This approach properly deals with the Malmquist bias as a 
stochastic (variable for clusters of the same observed mass) 
and uncertain correction
instead of a deterministic and perfectly
known constant according to the Planck and our baseline analyses 
\footnote{
The code used for the fitting is, after minor editing, the same Bayesian code
as described in and distributed by Andreon \& Berg\'e (2012), and used
in Andreon \& Condon (2014) for analysing the 
determination of the evolution of the richness--mass
scaling. 
We assume weak priors on slope, intercept,
and intrinsic scatter and solve for all variables at
once. We account, as mentioned, for the scatter, evolution, mass, and
selection function. We used a computationally
inexpensive 10 million long MCMC chain, discarding the initial 
10 thousand elements used for burn--in. By running
multiple chains, we checked that convergence is already achieved with  
short chains.}. 
We found $s=1.50\pm0.07$ (vs $s=1.51\pm0.07$ in the baseline analysis)
and $2.1\pm0.4$ (vs $\gamma=2.4\pm0.5$ ). The value $\gamma=2/3$ is rejected at
99.9 \%, showing that our results are robust to the precise
treatment of the mass+selection function.

\begin{figure}
\centerline{
\psfig{figure=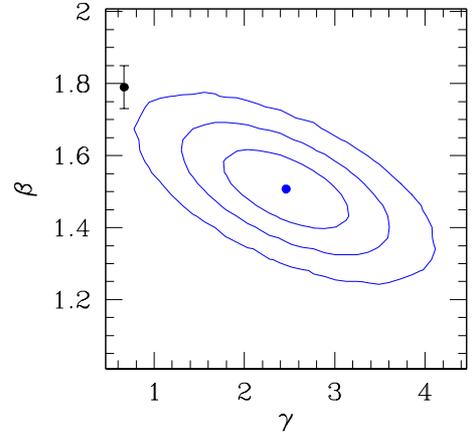,height=6truecm,clip=}
}
\caption[h]{Bounds on the mass slope $s$ and evolutionary
parameter $\gamma$ of the $Y_{ZS}$-mass relation (68, 95, 99.7\%, from inner to
outer contours). The point with error
bar marks the slope derived in Ade et al. (2014)
for the assumed $\gamma=2/3$. 
}
\end{figure}

\begin{figure}
\centerline{
\psfig{figure=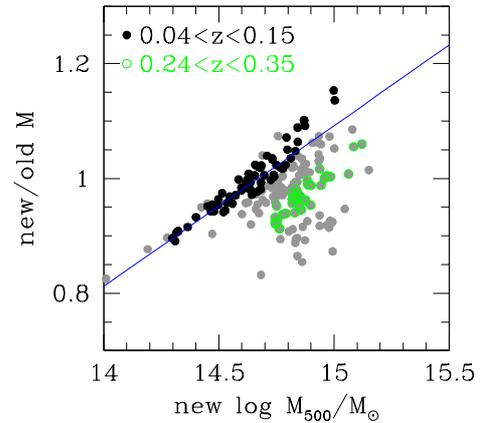,height=6truecm,clip=}
}
\caption[h]{Ratio between the recalibrated and original mass
from Planck $Y_{SZ}$ for the cosmological Planck sample (gray background)
and
for two redshift subsamples (black closed points and green open points).
The blue line has a slope equal to $0.28$.
}
\end{figure}

\section{Summary}

We refitted the values of the 71 Planck clusters in the
Planck calibration sample by allowing
evolution to be determined from the data, i.e. letting
the power of the $E(z)$ term be free, instead of fixed
to $2/3$. We used the very same input numbers
as the Planck team in order to emphasize that differences in
the results come from differences in the assumptions. 
We also repeated the analysis twice, once adopting the Planck team
corrections for Malmquist bias and once directly dealing with the mass
and selection function. The two analyses consistently
found (eq.~2) a shallower
$Y_{SZ}$--mass relation with a stronger evolution than assumed by the 
Planck team, i.e. a mass--dependent bias. 
The assumed (by the Planck team) self--similar evolution,
i.e. $\gamma=2/3$, is rejected at
$\ge 99.9$\% confidence by their own fitted data, which instead favour an
$E^{2.5\pm0.4}(z)$ evolution. 
This suggests caution in interpreting results
of works that assume a self--similar evolution and
determine, or revisit, the Planck proxy--mass
calibration under such an assumption because what is
said to be a mass bias could be a neglected evolutionary term.

Our revised calibration
is, at first sight, not useful for reducing the tension between the CMB--
and cluster--based cosmologies because at face value it increases, instead
of reducing, the already too large number of expected clusters for the Planck 
CMB cosmology.
However, this conclusion should be re--evaluated after
rederiving $Y_{SZ}$ proxy values of the cosmological sample
with 
a proxy--mass relation
with an evolution consistent with the data (i.e. with eq.~2).

The bottom line is that the non--self--similar evolution of the SZ proxy
cannot be ignored:  
$\gamma=2.5$ (supported by the data) has to be preferred to $\gamma=2/3$
(self--similar). 
Waiting for a re--derivation of the $Y_{SZ}$ values,
we suggest estimating the masses for Planck clusters using our eq.~2 and 
the $Y_{SZ}$ values listed in the Planck catalogue.

\begin{acknowledgements}
I thank the referee for suggesting the swap of the baseline and secondary
analysis to improve clarity and Daniel Gruen for pointing out his paper 
to me. I'm most grateful to Barbara Sartoris for comments on an early
version of this draft and to the Planck team for making 
the data table used for the $Y_{SZ}$-mass calibration publicly 
available.
I acknowledge 
Mauro Sereno for enlightening conversations on the subject of
this paper.
\end{acknowledgements}

{}

\end{document}